# Biomarker Integration and Biosensor Technologies Enabling AI-Driven Insights into Biological Aging

Jared A Kushner[1,2,3†], Mohit Pandey[1,4†]*, Sandeep (Sonny) S Kohli[1,5]

[1]*Diagen AI, Vancouver, British Columbia, Canada*

[2]*Department of Physics and Astronomy, University of Victoria*

[3]*Department of Biochemistry and Microbiology, University of Victoria*

[4]*University of British Columbia, Vancouver, British Columbia, Canada*

[5]*Intensive Care and Internal Medicine, Oakville Trafalgar Hospital, McMaster University, Oakville, ON, Canada*

*Corresponding author: mohit@diagen.ai

†These authors contributed equally

**Abstract**

As the global population continues to age, there is an increasing demand for ways to accurately quantify the biological processes underlying aging. Biological age, unlike chronological age, reflects an individual's physiological state, offering a more accurate measure of health-span and age-related decline. Aging is a complex, multisystem process involving molecular, cellular, and environmental factors and can be quantified using various biophysical and biochemical markers. This review focuses on four key biochemical markers that have recently been identified by experts as important outcome measures in longevity-promoting interventions: C-Reactive Protein, Insulin like Growth Factor-1, Interleukin-6, and Growth Differentiation Factor-15. With the use of Artificial Intelligence, the analysis and integration of these biomarkers can be significantly enhanced, enabling the identification of complex binding patterns and improving predictive accuracy for biological age estimation and age-related disease risk stratification. Artificial intelligence-driven methods including machine learning, deep learning, and multimodal data integration, facilitate the interpretation of high dimensional datasets and support the development of widely accessible, data-informed tools for health monitoring and disease risk assessment. This paves the way for a future medical system, enabling more personalized and accessible care, offering deeper, data-driven insights into individual health trajectories, risk profiles, and treatment response. The review additionally highlights the key challenges, and future directions for the implementation of artificial intelligence-driven methods in precision aging frameworks.

## 1. Introduction

While average life expectancy has steadily increased over recent decades, average health-span has remained relatively unchanged, with many individuals experiencing significant health challenges later in life [1,2]. This growing disparity has shifted the focus of the biomedical community towards not only increasing lifespan, but also maximizing the years spent in good health. Biological age, unlike chronological age, reflects the underlying state of the body's systems and has shown greater utility in predicting disease risk, functional ability, and overall health outcomes [3].

Biological aging is a complex, multi-system process driven by changes at molecular, cellular, and tissue levels. Twelve major categories have been identified as the "hallmarks of aging" which include genomic instability, telomere attrition, epigenetic alterations, loss of proteostasis, disabled macro autophagy, deregulated nutrient sensing, mitochondrial dysfunction, stem cell exhaustion, altered intercellular communication, dysbiosis, and chronic inflammation[4]. Even though the hallmarks of aging represent conceptual frameworks, they reflect underlying disruptions in key biological pathways. These physiological changes, while complex, can be quantified through measurable biomarkers that capture the molecular signatures of age-associated processes.

Aging can be assessed through a wide range of physiological, cognitive, and composite measures that reflect system-level decline. Physical markers such as grip strength, speed, gait, and balance control have been consistently associated with mortality, frailty, and loss of independence[5]. Cognitive functions including reaction time and executive control have also been used as a proxy for biological aging, particularly in the context of age-related neurodegeneration[6]. Cardiovascular markers such as pulse wave velocity and heart rate variability provide insight into vascular aging and CVD risk[7]. Integrated approaches like epigenetic clocks, frailty indices, and AI-driven composite scores combine multiple biological signals into predictive models that estimate biological age and forecast health trajectories [8,9]. Together, these diverse biomarker modalities offer comprehensive insights for monitoring the aging process across molecular and functional dimensions.

Among the large range of biomarkers, four have received particular attention due to their reproducibility, accessibility, and links to core aging pathways. These include C-reactive protein (CRP), interleukin 6 (IL-6), insulin like growth factor 1 (IGF-1), and growth differentiation factor 15 (GDF-15). These markers are all involved in key biological processes like inflammation, metabolic regulation, cellular stress, and proliferation; and can be used to estimate biological age and predict related disease risk [10].

Despite their clinical value, the analysis of biomarker data is often limited by the volume and complexity of information required to draw meaningful conclusions. Advances in artificial intelligence (AI), particularly machine learning (ML) and deep learning (DL), have enabled the efficient analysis of complex, high-dimensional biological data [11]. These techniques are now widely used to construct biological age clocks, enhance diagnostic accuracy, and predict health outcomes and disease risk with greater precision. Some recent examples include hematology-based bioclocks, and deep learning algorithms trained on multimodal datasets [12].

Beyond AI, advances in wearable biosensors now allow for continuous, non-invasive monitoring of physiological signals, providing real time data feedback for health assessment. When integrated with AI, this continuous data stream enables consistent and accurate prediction of disease risk and overall biological age, offering valuable insights into an individual's current and future health trajectory [13]. This convergence of biomarkers, biosensors, and AI holds promise for a new era of personalized aging diagnostics and preventative medicine [11]; [14].

This review synthesizes current knowledge on key biochemical markers of aging and explores how advances in biosensor technology and AI are transforming their measurement, interpretation, and clinical applications.

## 2. Biomarkers of Aging

Biomarkers of aging are measurable indicators that reflect the biological processes underlying aging, offering critical insights into molecular, cellular, and physiological mechanisms. They provide a quantification of age-related decline, offering a measurable distinction between biological age and chronological age. This distinction is critical for assessing healthspan, predicting disease risk, and monitoring the efficacy of therapeutic interventions [15].

Age-related biomarkers fall broadly into two categories: biochemical which include a person's metabolites and circulating proteins; or phenotypical such as one's gait and grip strength. Biochemical markers are intrinsic to biological processes and are tightly linked to pathways in chronic inflammation, cellular senescence, and metabolic regulation [16]; [17]. Phenotypical biomarkers, in contrast, assess physical function, and serves as a proxy for musculoskeletal integrity and neuromuscular function [18].

An ideal biomarker of aging should be biologically relevant, reproducible, and accessible through non-invasive techniques [19].

### 2.1 Applications of Biomarkers in Aging

Biomarkers of aging are increasingly applied in research and clinical settings to estimate biological age, predict health outcomes, and evaluate intervention efficacy. Composite biomarker-based indices, such as the Dynamic Organism State Indicator (DOSI) and the Physiological Frailty Index, have demonstrated improved performance over chronological age in forecasting all-cause mortality, hospitalization, and frailty [20]; [21]. These models integrate routinely measured blood and physiological variables, providing scalable tools for population-level aging surveillance and personalized risk assessment. In clinical trials, biochemical markers such as GDF-15, IL-6, IGF-1, and CRP are increasingly adopted as surrogate endpoints to track biological responses to interventions including dietary restriction, exercise, and pharmacologic agents [22]; [17]; [23]. Longitudinal changes in these markers can reflect alterations in inflammatory burden, metabolic stress, or resilience, offering early signs of intervention effectiveness. As wearable biosensors and remote sampling technologies improve, continuous biomarker monitoring may allow dynamic tracking of individual health, enabling adaptive and precision-based interventions across the aging spectrum. Furthermore, the alignment of biomarkers with known aging pathways, such as inflammation, metabolic regulation, and cellular stress responses, strengthens their utility in assessing the complex multifactorial nature of aging.

Each of the twelve hallmarks of aging are driven by different combinations of biological pathways, each with distinctive associated biochemical markers. The four biochemical markers previously mentioned (GDF-15, IL-6, IGF-1, and CRP) form a strong base for age related diagnostics due to their direct involvement in aging mechanisms. Collectively, these biomarkers provide broad coverage across all twelve hallmarks, with each hallmark intersecting with at least one of the four as seen in figure 1. This comprehensive overlap underscores their value as an integrated panel for monitoring the complex biology of aging.

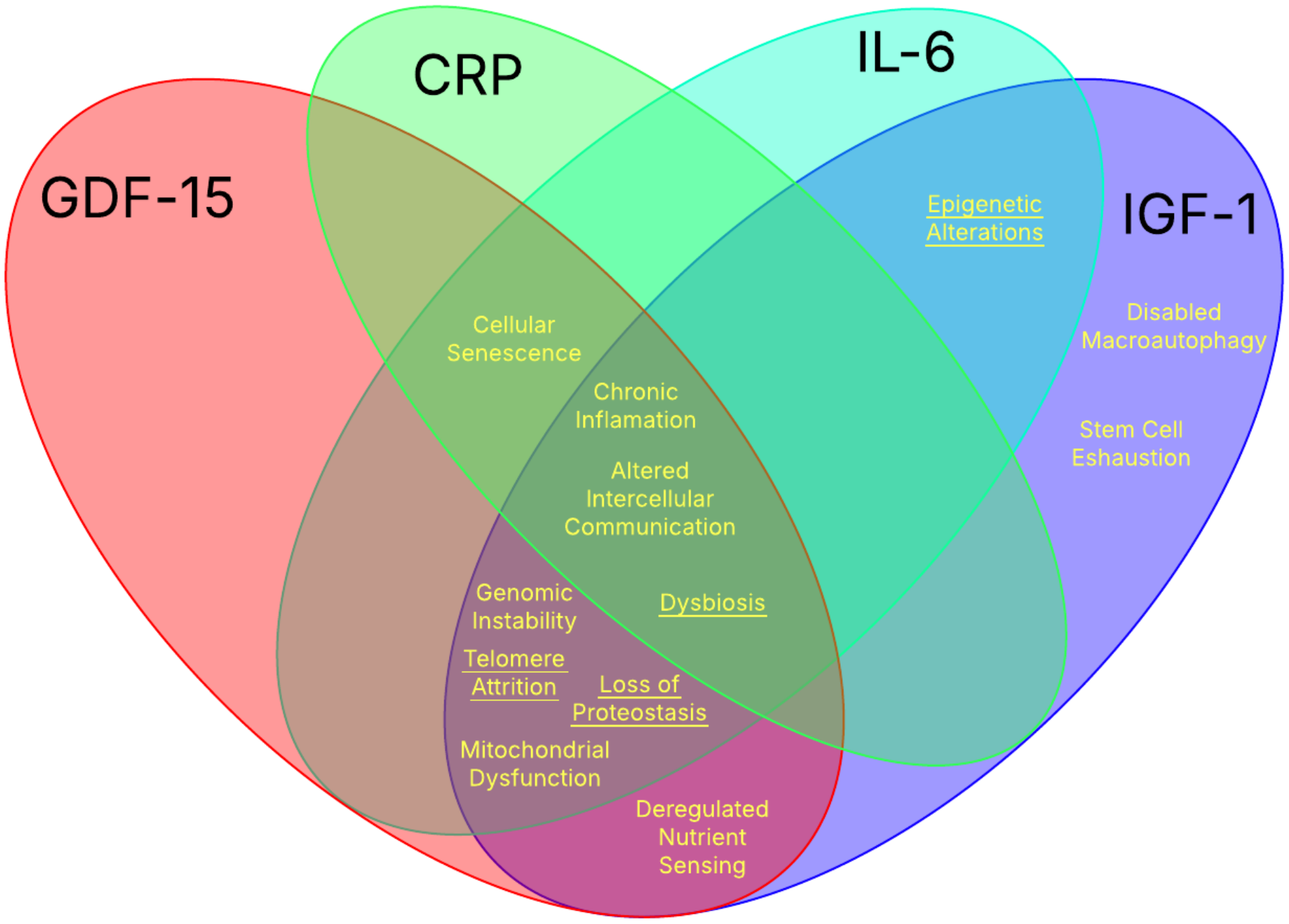

**Figure 1.** The intersection of the twelve hallmarks of aging with the four key biochemical biomarkers. The Venn diagram illustrates the intersection between the established hallmarks of aging and the biological processes influenced by the four well-studied biochemical aging biomarkers: CRP, IL-6, IGF-1, and GDF-15. Each biomarker contributes to distinct and overlapping hallmarks indicated in yellow (underlined hallmarks signify indirect associations).

## 2.2 Biomarkers in Related Pathways

Several major biological pathways are involved in the aging process, offering insight into potential targets for biomarkers and therapeutics. The insulin/IGF-1 signaling pathway plays a central role in lifespan modulation and has been extensively linked to longevity across species [24]. The MAPK/ERK pathway regulates cellular proliferation and responses to oxidative stress, contributing to age-related cellular damage. Similarly, the PI3K/Akt/mTOR pathway is a major regulator of nutrient sensing and growth, and its dysregulation has been associated with age-related metabolic decline and reduced autophagy [25]; [26]. Chronic low-grade inflammation, "inflammaging", is another force of aging, mediated by pro-inflammatory cytokines such as IL-6 and CRP, which are involved in immune signaling and systemic inflammatory responses [27,28]. These inflammatory markers reflect the broader phenomenon of immunosenescence, contributing to increased vulnerability to infections, cancer, and autoimmune diseases. Table 1. highlights the pathways and functions related to each of the four key biochemical markers along with their respective clinical associations. Together, these pathways form the molecular foundation for the understanding of aging and identifying relevant biomarkers.

## 2.3 The Key Biochemical Markers of Aging

*C-Reactive Protein*

CRP is a highly conserved member of the pentraxin family critically involved in innate immunity. In its native conformation, CRP has a homopentameric structure composed of five identical non-glycosylated subunits arranged symmetrically around a central pore [29]. In addition to the pentameric isoform (pCRP), a monomeric form of CRP (mCRP) is generated through irreversible dissociation at sites of inflammation [30]. These two conformations play respective roles in the body's immune system. pCRP is found primarily circulating in blood (serum) and functions as a pattern recognition molecule facilitating opsonization and the activation of the complement pathway [31]. On the other hand, mCRP is typically tissue bound and exhibits potent pro-inflammatory properties through leukocyte recruitment and enhanced cytokine production [32]. In response to foreign materials or damaged tissues, circulating CRP (pCRP) can bind phosphocholine on the cell's surface via calcium-dependent ligand binding leading to the initiation of the classical complement pathway [33]. The activation of this pathway is mediated through the recruitment of C1q and the direct binding to the Fc region on IgG and IgM antibodies (immunoglobulin) [29]. Additionally, CRP can engage with Fcϒ receptors (notably FcϒRI and FcϒRII), promoting immune cell activation and cytokine production [34]. With the use of high sensitivity assays, circulating CRP levels can be measured for insights into disease risk and inflammatory status. Given its dual role in immune surveillance and inflammation, CRP, particularly high-sensitivity CRP (hsCRP), serves as a robust biomarker of aging. Elevated levels of hsCRP have been consistently linked with age-related diseases such as CVD, frailty and cognitive decline [35].

*Insulin-like Growth Factor-1*

IGF-1 plays a central role in anabolic signaling, cellular growth, and development, making it a key mediator in the aging process. IGF-1 is a key regulator of the insulin/IGF-1 signaling pathway (IIS), binding directly to the IGF-1 receptor (IGF-1R), which modulates lifespan and cellular growth through downstream cascades like the PI3K/Akt/mTOR and MAPK/ERK pathways [36].

While IGF-1 has been proven to play a role in lifespan and longevity, its exact role in aging is still hard to understand. While low IGF-1 signaling is linked to increased longevity in model organisms, its role in humans is more complex. Both high and low levels of circulating IGF-1 are associated with increased risk of morbidity and mortality, suggesting a U-shaped relationship [37]. Elevated levels have been linked to cancer, while low levels are associated with frailty and CVD [38]. Given IGF-1's strong association with key aging pathways and diseases, highlights its utility as a biomarker of biological age.

*Interleukin-6*

IL-6 is a multifunctional pro-inflammatory cytokine that plays a critical role in the body's immune response and has been continuously implicated in the biology of aging. Produced mainly by immune cells and hepatocytes in the liver, IL-6 initiates the acute-phase response by stimulating the production of CRP [39]. IL-6 levels increase with age [40,41], contributing to a persistent low-grade inflammatory state known as "inflammaging", which is associated with numerous age-related diseases such as CVD, frailty, sarcopenia and neurodegeneration [42,43]. Elevated IL-6 has been consistently linked to physical decline and increased mortality risk in older adults [44]. In particular, IL-6 impairs muscle regeneration with age by disrupting satellite cell function and promoting catabolic pathways, thereby contributing to sarcopenia and reduced mobility [45]. Furthermore, chronic IL-6 elevation exacerbates oxidative stress and mitochondrial dysfunction, reinforcing the cycle of tissue damage and systemic inflammation [46]. Given its central role in mediating age-related inflammatory responses and predicting adverse health outcomes, IL-6 is considered a key biomarker of biological aging.

*Growth differentiation factor-15*

GDF-15 is a stress-responsive cytokine in the transforming growth factor-beta (TGF-β) family. It is widely recognized as a biomarker of mitochondrial dysfunction [47], cellular stress [48], and tissue injury [49], with levels increasing in response to oxidative metabolic stress. GDF-15 has been strongly associated with aging and age-related pathologies, including CVD, cancer, and frailty, showing consistently elevated levels in older adults [50]. Furthermore, GDF-15 is increasingly recognized as a biomarker intricately associated with epigenetic aging. Epigenome-wide association studies have identified specific CpG methylation sites that are associated with GDF-15 expression, suggesting a potential regulatory relationship between GDF-15 and age-related epigenetic alterations [51]. Additionally, Recent studies have shown that circulating GDF-15 correlates strongly with several DNA methylation-based clocks, such as GrimAge, PhenoAge, Hannum, and Zhang [52], thus proving its value as a biomarker of aging. While GDF-15 correlates with multiple epigenetic clocks, its specificity remains questionable due to its elevation in diverse non-age-related acute stress states including cancer, CVD, and renal disease [50,53].

**2.4 Additional Key Biochemical Markers**

Biological aging is often marked by the dysregulation of cell cycle and tissue maintenance pathways, where specific protein biomarkers such as p21, p16, and Klotho have demonstrated significant relevance. The proteins p21 and p16 are cyclin-dependent kinase inhibitors upregulated during cellular senescence. The accumulation of these kinase inhibitors reflects DNA damage responses and irreversible growth arrest in aging tissues [54]. Notably, p16 is considered one of the most robust markers of senescent cell burden in vivo and increases predictably with age across multiple tissues [55]. Similarly, p21 serves as an early senescence mediator and is closely associated with stress-induced and telomere-dependent senescence pathways [56]. In contrast, the Klotho protein functions as a longevity hormone, primarily through its role in suppressing oxidative stress and modulating IGF-1 signaling. Klotho expression declines with age and its deficiency leads to accelerated aging phenotypes in mice, whereas its overexpression extends lifespan [57,58]. These proteins not only act as mechanistic biomarkers of cellular aging but also represent therapeutic targets in regenerative medicine and senescence-modulating interventions.

There are hundreds of biomarkers to choose from when trying to estimate one's biological age. For a broader set of aging biomarkers, including their detection methods and applications, see Table 2.

## 3. Bioclocks

Biological clocks are computational models that estimate an individual's biological age by analyzing patterns in molecular, physiological, and functional biomarkers. These clocks integrate input from multiple data sources, including DNA methylation profiles, blood-based biomarkers such as CRP, IGF-1, IL-6, and GDF-15, as well as physical and cognitive performance measures. The resulting age estimates often outperform chronological age in predicting morbidity, mortality, and overall health outcomes [59,60]. Epigenetic clocks (e.g., Horvath, GrimAge, PhenoAge) have pioneered this field, using machine learning models trained on large datasets of methylation signatures [8,61]. A selection of currently available biological age estimation tools, along with their features and data modalities, is summarized in Table 3. Recent approaches to biological clocks incorporate multimodal data sources. These include real-time signals from wearable biosensors such as heart rate variability, physical activity levels, and sweat-based biomarkers. This integration improves both the granularity and temporal resolution of age estimation models.

AI plays a critical role in refining these models, especially through deep learning and ensemble techniques that can learn nonlinear relationships between inputs and outcomes. Advanced models can continuously update biological age estimates as new data are collected, enabling adaptive biological clocks that reflect acute changes (e.g., during illness or recovery) and chronic trends (e.g., long-term aging rate). Biosensors provide the necessary infrastructure for continuous data acquisition, allowing biological clocks to function in real time and outside of clinical settings. This convergence enables dynamic health risk stratification, early warning systems for age-related diseases, and longitudinal monitoring of therapeutic interventions, all within a personalized framework [14,62,63]. Overall, the combination of AI, biosensors, and biomarker data has the potential to revolutionize health care and longevity research.

## 4. Wearable Biosensors and Real-Time Monitoring

Recent advances in biosensing technologies have enabled continuous, real-time monitoring of age-related biomarkers, marking a significant shift from static measurements to dynamic health tracking. This shift paves the way for personalized health care, providing real-time feedback for disease progression and overall health trajectory. Wearable biosensors, especially those capable of non-invasive data collection, are now a critical tool in aging research and personalized health management. Traditional biosensors rely on biochemical detection methods such as electrochemical, optical, or piezoelectric sensing to quantify circulating analytes. These analytes can include a large range of biomarkers; however, this review is focused solely on age-related biomarkers.

Developments in biosensing have expanded the capabilities, allowing for the detection of biomarkers in a wide range of biological mediums such as sweat, saliva, and interstitial fluid. An innovation in biosensing is now able to monitor hsCRP levels in sweat via a wearable wireless patch, therefore bypassing the blood-based assays typically used for inflammatory markers [13]. Additionally, wearable microneedle patches hold significant promise for the real-time monitoring of biomarkers related to chronic disease and aging by enabling the non-invasive sampling of interstitial fluid [64]. These patches utilize microneedle arrays that penetrate the stratum corneum without reaching nerve-rich regions, minimizing pain while allowing for sustained sampling [65]. Other studies have validated the feasibility of using wearable patches to monitor cytokines and protein biomarkers in interstitial fluid or sweat, supporting their application in aging research and chronic disease surveillance [66–68]. The potential clinical utility of such patches is significant, particularly in early disease detection and management of chronic disease, inflammation, and aging.

## 5. Modeling Biological Aging with Machine Learning

ML methods especially with the recent advancements in generative modeling have opened new directions for understanding and forecasting biological aging. Generative models simulate future physiological states, capture latent biological signatures, and enable hypothetical counterfactual analysis. In parallel, predictive ML methods have made significant advances in developing accessible, interpretable, and often clinically actionable biomarkers of aging. Grouping these advances by their application domains provides a clearer picture of how diverse modeling approaches are reshaping aging research.

### 5.1. **Brain Aging and Neurodegeneration**

One key frontier is modeling brain aging, particularly using neuroimaging. Conditional generative models trained on cross-sectional 18F-FDG PET scans have been shown to simulate future metabolic topographies in cognitively normal individuals [69]. These models reveal genotype-specific trajectories; for example, APOE4 carriers show early decline in regions implicated in Alzheimer's pathology.

Building on this, the SynthBrainGrow model [70] uses diffusion-based methods to generate realistic MRI scans showing progressive structural changes like cortical thinning and ventricular enlargement. These simulations have been validated against actual follow-up scans, demonstrating their utility in augmenting datasets for neurodegeneration studies.

Conditional normalizing flow-based approaches learn bidirectional mappings between morphology and chronological age [71] through MR images, enabling both brain age predictions and age-conditioned reconstructions of brain anatomy to represent aging trends.

### **5.2. Epigenetic and Molecular Aging Clocks**

A second major application is predicting biological age from high-dimensional omics data. Variational autoencoders trained on genome-wide DNA methylation (DNAm) profiles can reconstruct age while discovering biologically interpretable latent features linked to aging [72]. Chromosome-wise autoencoders [73] further enhance compression while identifying regulatory CpG sites associated with age acceleration. Cytosine-phosphate-guanine (CpG) sites are regions of DNA where a cytosine nucleotide is bonded to a guanine through a phosphate bond and serves as hotspots for DNA methylation.

Similarly, DL methods have outperformed traditional linear models in age prediction by capturing nonlinear interactions among CpG sites [74]. Due to the non-linear modeling capacity of the neural networks, they offer higher generalizability across tissues and show heightened sensitivity to aging-related diseases, including multiple sclerosis and diabetes.

In male-specific epigenetic clock, a support vector machine trained on Y-chromosome CpG sites created the first sex-specific methylation clock, achieving strong age correlation [75]. Similarly, a cardiovascular health study (CHS)-specific epigenetic clock based on five age-related genes was trained on targeted CpG data, with random forest regression outperforming other models [76].

A recent study [77] has further shown that even biomechanical data like upper-extremity function can be used to classify frailty status. By combining movement velocity and muscle co-contraction metrics with LSTM networks, the model effectively distinguished between frail and non-frail older adults.

### 5.3. Simulating Aging Trajectories and Forecasting Longevity

Simulating the future course of aging is essential for both basic research and intervention studies. In mice, frailty indices (FIs) have been used to train two random forest models: one predicts chronological age (FRIGHT), while the other (AFFRAID) estimates life expectancy [78]. These models have also proven effective in detecting the benefits of longevity interventions, such as drugs or gene modifications, well in advance.

On the human side, the DJIN model [79] represents a stochastic dynamical system that maps how health variables interact over time. Trained on the English Longitudinal Study of Aging, DJIN predicts health trajectories and survival outcomes while uncovering directed relationships among physiological and functional indicators. Unlike traditional survival models, it emphasizes interpretability and interaction structure.

Meanwhile, Sundial [80] offers a generative diffusion-based framework that models the molecular aging dynamics (e.g., from transcriptomic, methylation, or other omics data) using a diffusion field without relying on chronological age. It reconstructs personal "aging roadmaps" from cross-sectional omics, enabling the identification of individuals aging faster than average which is a valuable trait for early intervention trials.

### 5.4. Aging Biomarkers from Lifestyle, Microbiome, and Biochemistry

Beyond molecular data, several models have exploited accessible lifestyle, microbial, and biochemical information to predict aging.

A recent study in cerebrovascular disease patients [81] used vascular risk factors, organ damage scores, and habits to predict age acceleration. Multilayer perceptrons and elastic net models yielded the best results, showing that such non-invasive data moderately capture epigenetic aging. The integration of microbiome and metabolome data offers another promising avenue. A study on 568 healthy individuals [82] trained XGBoost models using 16S rRNA gene sequencing of fecal samples and urine metabolite profiles. Richer microbiome diversity and specific bacterial genera were associated with age, and combining omics improved prediction accuracy, highlighting the utility of non-invasive biomarker fusion.

Another effort built a physiological age model that is independent of chronological age, with biochemical and physiological features such as high-density lipoprotein, pulse wave velocity, and psychological traits [83] using statistical ML algorithms. The resulting physiological aging rate correlated strongly with real age and mortality risk and showed ~30% heritability, suggesting potential for future genetic studies.

Finally, predictive frameworks that integrate gene expression and protein–protein interaction networks have shown that modeling dynamic, weighted interactions lead to better identification of aging-associated genes [84]. These models outperform static subnetworks and may guide new longevity drug targets.

### 5.5. Identity-Preserving Face Aging and Biometric Analysis

Generative models also contribute to non-invasive aging tracking using facial features. One approach uses diffusion autoencoders with text-guided embeddings to simulate diverse aging trajectories from a single face image and textual prompts (e.g., "old scientist", "90s fashion") [85]. These can be used for health assessments, biometric verification, and age estimation under various lifestyle scenarios. Other methods leverage invertible neural networks to disentangle age from identity, preserving demographic features while generating forward or reverse aged faces [86]. These models offer robustness and realism often missing in GAN-based face aging methods.

### 5.6 AI-integrated Biosensors

The integration of AI into biosensor platforms allows for high-throughput data analysis, enabling the identification of health patterns for biological age estimation. DL models trained on step count and wearable data have already been able to predict morbidity risk with accuracy [20], demonstrating the implementation of AI-integrated biosensors. Similarly, AI-enhanced biosensing platforms are increasingly used at point of care to process complex bio-signal data quickly and accurately, enabling personalized interventions [87]

Beyond data analysis, AI is now being applied upstream in the design of the biosensors themselves, particularly in the development of molecular binders. AI models, including deep generative architectures and reinforcement learning algorithms, can be used to design high-affinity protein binders for specific analytes such as cytokines, metabolites, or other age-related biomarkers. In the case of CRP, IL-6, GDF-15, and IGF-1, a comprehensive set of validated and computationally predicted binders, along with their sequences and structural representations is detailed in Table 4. These computational approaches enable the de novo design of synthetic receptors with tailored binding kinetics, selectivity, and stability, which are critical for the manufacturing of accurate, high precision biosensors. Recent studies have demonstrated that deep learning guided diffusion models and protein design frameworks like RFdiffusion can successfully generate binders for challenging targets [88]. This reduces the dependence on wet-lab experimental screening and accelerates the creation of custom binders for biosensor manufacturing.

## 6. Challenges

Despite the promise of integrating AI, biosensors, and biomarkers in biological age estimation, several critical challenges must be addressed to ensure accuracy, equity, and real-world applicability. One major concern is algorithmic bias, which arises when training datasets do not represent the diversity of global populations. AI models built on narrow demographic or clinical data can produce skewed or inaccurate results, particularly for underrepresented groups [89]. Developing models that generalize well requires large, diverse datasets that capture a wide range of physiological, genetic, and environmental variability.

Privacy and security are central to the ethical implementation of large-scale data sets containing confidential health information. This requires robust data protection measures and compliance with privacy regulations such as HIPAA and GDPR. Without secure data infrastructures and transparent consent mechanisms, public trust and clinical adaptation may be compromised [90].

The cost of wearable biosensors along with the computational requirements to run complex AI-based platforms is a potential limiting factor for widespread adoption. Ensuring that the technology is user friendly, accessible, and non-invasive is essential for its utilization in both clinical and everyday settings.

## 7. Future Directions

Emerging developments in AI and biosensor integration are poised to further advance the field of aging research. First, the implementation of multimodal aging clocks that synthesize data from molecular, physiological, and behavioral domains is expected to improve the accuracy of biological age estimates. These models could incorporate additional data streams, such as sleep architecture, microbiome composition, or metabolomics, increasing their diagnostic utility.

AI-guided biosensor design is expected to become more sophisticated, incorporating reinforcement learning and structure-aware generative models to design highly specific protein binders for age-related targets. Recent work has already demonstrated the feasibility of using language-based generative models to design protein structures that bind previously undruggable targets [91]. Additionally, the adoption of federated learning could facilitate model training across decentralized datasets, enabling robust and generalizable predictions while preserving user privacy [92]. In the commercial sphere, integration into existing consumer health devices could democratize access, providing individuals with feedback on aging trajectories and enabling personalized, non-invasive health interventions.

As these systems mature, interdisciplinary collaborations across geroscience, biomedical engineering, and clinical practice will be critical to validating models, improving interpretability, and ensuring regulatory compliance. These efforts will ultimately shape the development of scalable, ethical, and clinically relevant tools for biological aging assessment.

## 8. Conclusion

Biomarkers such as CRP, IL-6, IGF-1, and GDF-15 provide critical insight into the physiological underpinnings of aging, especially when measured dynamically with modern biosensors and interpreted through AI models. When used in combination, these technologies enable a more refined understanding of health trajectories, allowing researchers and clinicians to track aging in real time and adapt interventions accordingly. However, meaningful application of these tools must address persistent limitations. These include algorithmic bias, limited data diversity, high development costs, and accessibility barriers. Moreover, standardization of methods, cross-disciplinary validation, and patient compliance remain key obstacles to clinical integration. Despite these hurdles, the convergence of biomarkers, biosensors, and AI continues to hold substantial promise. The field is moving steadily toward personalized, data-driven health management, supporting longer and healthier lives through precision aging diagnostics and targeted interventions.

| Biomarker | Pathways | Primary Function | Clinical Associations |
|---|---|---|---|
| CRP | Classical Complement Activation, FcγR signaling | Inflammation Marker, Acute phase reactant | CVD, Diabetes, Cancer, Autoimmune |
| IL-6 | JAK-STAT, Ras-MAPK, PI3K-Akt | Pro-inflammatory cytokine | Frailty, Chronic Inflammation, neurodegeneration |
| IGF-1 | Insulin/IGF-1 signaling, PI3k/Akt, mTOR | Growth Factor, Regulates Metabolism | Longevity, Sarcopenia Metabolic Syndrome, Cancer |
| GDF-15 | TGF-β, Mitochondrial stress response | Stress-responsive Cytokine, Marker of Mitochondrial Dysfunction | Multimorbidity, Mortality, Cachexia, CVD |

**Table 1.** Summary of key age-associated biomarkers, their molecular binders, associated pathways, biological functions, and clinical relevance.

| Marker | Category | Domain | Measurement Method(s) | Applications |
|---|---|---|---|---|
| **Heart Rate Variability** | Phenotypic | Vascular | ECG, Wearable Heart Rate Sensors | [93] |
| **Hair colour (gray)** | Phenotypic | Integumentary | Visual inspection | [94] |
| **gait** | Phenotypic | musculoskeletal | Motion capture, Wearable Sensors | [95] |
| **speed** | Phenotypic | musculoskeletal | Time-Based Tests | [96] |
| **Grip strength** | Phenotypic | musculoskeletal | Hand Dynamometer | [97] |
| **DNA methylation** | biochemical | Genetic | Illumina Arrays, Bisulfite Sequencing | [98] |
| **Insulin** | biochemical | Endocrine | Blood Assay ELISA | [99] |
| **Tumor necrosis Factor (TNF-alpha)** | biochemical | Endocrine | ELISA, Multiplexed Cytokine Assays | [100] |
| **LDL (oxLDL)** | Biochemical | Vascular | Blood Test, ELISA | [101] |
| **Telomere Length** | Biochemical | Genetic | qPCR, Southern Blot, Flow-FISH | [102] |
| **Leptin** | Biochemical | Endocrine | Blood Test, ELISA | [103] |
| **Adiponectin** | Biochemical | Endocrine | Blood Test, ELISA | [104] |
| **Myeloperoxidase** | Biochemical | Immune | Blood Test, Immunoassays | [105] |
| **Chair rise time** | Phenotypic | Musculoskeletal | Timed chair stand test, 5-time chair rise test (5CRT) | [106] |
| **Balance time** | Phenotypic | Musculoskeletal | Timed balance test | [107] |
| **Reaction time** | Phenotypic | Cognitive | Computerized cognitive tests, psychometric tools | [108] |
| **Sleep fragmentation** | Phenotypic | Cognitive | Actigraphy, polysomnography, wearable trackers | [109] |
| **Pulse wave Velocity** | Phenotypic | Vascular | Applanation tonometry, oscillometric devices | [110] |
| **Frailty index** | Phenotypic | Overall | Comprehensive geriatric assessment (CGA), clinical scoring | [111] |
| **6-minute walk test** | Phenotypic | Musculoskeletal | Timed distance walk | [112] |
| **VO2 max** | Phenotypic | Vascular | Cardiopulmonary exercise testing (CPET) | [113] |
| **Bone mineral density** | Phenotypic | Musculoskeletal | Dual-energy X-ray absorptiometry (DEXA), QCT | [114] |
| **Skin elasticity** | Phenotypic | Integumentary | Cutometer®, elastography, indentation-based devices | [115] |
| **P16** | Biochemical | Cell Cycle | qPCR, Western blot, immunohistochemistry | [116] |
| **P21** | Biochemical | Cell Cycle | Western blot, ELISA, immunofluorescence | [117] |
| **Klotho Protein** | Biochemical | Endochrine | ELISA, immunoblotting, blood/urine assays | [118] |
| **HbA1C** | Biochemical | Endochrine | HPLC, point-of-care devices, immunoassays | [119] |

**Table 2.** In-depth summary table of known biomarkers of aging with measurement methods and applications (excluding the four key biomarkers).

| **Name** | **Type** | **Dataset** | **Description** | **References** |
|---|---|---|---|---|
| DNAmAge | Code Base | DNA Methylation | First generation epigenetic clock | [8] |
| TruAge | Webtool | DNA Methylation | Commercial epigenetic age testing webtool | [120] |
| ClockBase | Webtool/ Database | DNA Methylation, blood chemistry | Interactive platform for comparing biological clocks across datasets | [121] |
| BioAge | Code Base | Clinical Biomarkers | Mortality-based biological age estimator using clinical biomarkers | [122] |
| AltumAge | Code Base | DNA Methylation | Deep learning-based pan-tissue epigenetic clock implemented in TensorFlow and PyTorch | [74] |
| DeepMAge | Code Base | DNA Methylation | Deep neural network model for predicting biological age from DNA methylation data | [123] |
| MethylNet | Code Base | DNA Methylation | Modular deep learning framework for methylation data analysis, including age prediction | [124] |
| dnaMethyAge | Code Base | DNA Methylation | R package supporting multiple DNA methylation clocks for age prediction | [125] |
| methylclock | Code Base | DNA Methylation | Bioconductor package to estimate DNA methylation age using various established clocks | [126] |
| epigeneticclock | Code Base | DNA Methylation | calculates DNA methylation age using Horvath's method | [127] |
| meffonym | Code Base | DNA Methylation | R package for DNA methylation-based indices of exposure and phenotype, including age estimation | [128] |
| TallyAge | Webtool | DNA Methylation | commercial service for biological age estimation based on DNA methylation patterns obtained from cheek swab samples | [129] |

**Table 3.** Summary of Available Biological Age Estimation Tools and Platforms.

| Binder | Type | Structure | Affinity | Source | Pros & Cons |
|---|---|---|---|---|---|
| C-Rective Protein | | | | | |
| Phosphocholine | Small Molecule | PubChem CID: 1014<br><br>PDB: 1B09 (Complex) | $Kd \approx 4.8\ \mu M$ | [130] | High specificity for CRP, small size; not ideal for high-throughput sensing due to low signal generation alone |
| CRP DNA Aptamer | DNA Aptamer | 5′-CtAGTTCtGCCtTAATATGGtCGGtTAAGC… (48 nt; t = 5-(guanidino)-dU modification) | $Kd = 6.2\ pM$ | 131 | Pros: Ultra-high affinity (picomolar) CRP binder, even higher than original aptamer (53 pM); useful for ultrasensitive CRP detection.<br><br>Cons: Requires modified bases for affinity; potential structural complexity and stability issues in vivo; selected in vitro, not yet clinically tested. |
| 4-C25L22-DQ | Peptide - Small molecule | 4-C25L22-2-oxo-1,2-dihydroquinoline-8-carboxylic acid (DQ) | $Kd = 760\ nM$ | 132 | Pros: High affinity, calcium-independent, serum-selective, tunable scaffold, noncanonical CRP site.<br><br>Cons: Undisclosed sequence, complex synthesis, untested in vivo, unknown stability. |
| P3 | Peptide | VHWDFRQWWQPS | $Kd = 35 \pm 1.2\ nM$ | 133 | Pros: Strong docking, easy to synthesize<br><br>Cons: No SPR binding, lacks calcium site, unproven in serum |
| 4-C10L17PC6 | Peptide-small molecule | 4-C10L17-PCh6 | $Kd < 10\ nM$ | 134 | Pros: Strong affinity, irreversible binding, highly specific, tunable synthetic design<br><br>Cons: calcium-dependent, complex synthesis, no in vivo or structural data |
| 3-D10L17-PC6 | Peptide-small molecule | 3- D10L17 (sequence Ac-NAADJEARI KHLAERJK ARGPVD CAQJAEQL ARAFEAFA RAG-CONH2) | Low nanomolar dissociation constant | [135] | Pros: Strong ELISA binding, selective for CRP, multivalent, fully synthetic<br><br>Cons: No KD reported, calcium-dependent, no structural/in vivo data. |

| Insulin-like Growth Factor-1 | | | | | |
|---|---|---|---|---|---|
| IGF-1R | Glycoprotein | PDB 7YRR | Kd ≈ 0.16 $nM$ | [136] | Pros: Natural receptor, very high specificity, can be used in competitive assays<br><br>Cons: Large size makes it difficult to immobilize on sensors, Expensive to produce, |
| IGFBP-1 | Protein | UniProt ID: P08833 | K(a) range: 1x10^4 – 9x10^5 M^-1 s^-1<br><br>K(d) range: 1.5x10^-5 – 2x10^-4 s^-1 | [137] | Pros: High affinity, smaller size enables immobilization, extends IGF-1 half-life<br><br>Cons: Requires reducing agents to avoid disulfide aggregation, Binds IGF-2 with similar affinity |
| IGFBP-2 | Protein | UniProt ID: P18065 | K(a) range: 1x10^4 – 9x10^5 M^-1 s^-1<br><br>K(d) range: 1.5x10^-5 – 2x10^-4 s^-1 | [137] | Pros: High affinity, smaller size enables immobilization, extends IGF-1 half-life<br><br>Cons: Requires reducing agents to avoid disulfide aggregation, Binds |
| | | | | | IGF-2 with similar affinity |
| IGFBP-3 | Protein | UniProt ID: P17936 | K(a) range: 1x10^4 – 9x10^5 M^-1 s^-1<br><br>K(d) range: 1.5x10^-5 – 2x10^-4 s^-1 | [137] | Pros: Most abundant IGF binder in serum, high affinity<br><br>Cons: Large size, heavily glycosylated, equal affinity to IGF-2 |
| IGFBP-4 | Protein | UniProt ID: P22692 | K(a) range: 1x10^4 – 9x10^5 M^-1 s^-1<br><br>K(d) range: 1.5x10^-5 – 2x10^-4 s^-1 | [137] | Pros: high affinity, smaller functional fragments exist, blocks IGF-1R binding site<br><br>Cons: low serum abundance, binds to IGF-2 with similar affinity |
| IGFBP-5 | Protein | PDB 1H59 (Complex)<br><br>PDB 1BOE (Domain) | K(a) range: 1x10^4 – 9x10^5 M^-1 s^-1<br><br>K(d) range: 1.5x10^-5 – 2x10^-4 s^-1 | [137] | Pros: high affinity, mini-IGFBP-5 domain retains full binding site, Resistant to proteolysis in serum<br><br>Cons: moderately binds to IGF-2, multiple disulfides |
| IGF1-25R | DNA Aptamer | ATCCGTCA CACCTGCT CTGCAAGC ATTCATAT TGGTTGGT GGAAGTGG GGGGGGTG GT GTTGGCTC CCGTAT | High affinity; low nM range<br><br>Assay LOD: ~16ng/mL | [138] | Pros: high specificity, small size, rapid binding, reversible denaturation<br><br>Cons: Kd not explicitly measured, sensitive to nuclease degredation |
| IGF-1-F1-1 | Peptide | PDB 1LB7 | $IC50$=7.2±3.4 $\mu M$ | [139] | Pros: small size suitable for biosensor, competitively inhibits |
| | | | | | IGF-1's interactions, easy to synthesize<br><br>Cons: potential cross-reactivity with IGF-2, moderate affinity may require high concentrations for detection |

| Interleukin-6 | | | | | |
|---|---|---|---|---|---|
| Siltuxi-mab | Antibody | DrugBank ID: DB09036 | high-affinity ~nM-pM | [140] | Pros: FDA-approved for Castleman's disease; neutralizes IL-6 potently.<br><br>Cons: Large biologic (IV infusion); risk of immunogenicity and infections due to immune suppression. |
| Siruku-mab | Antibody | KEGG Entry: D10080 | Kd ≈ 0.175 pM | [141] | Pros: High specificity IL-6 neutralization; showed efficacy in rheumatoid arthritis trials<br><br>Cons: Not approved (development halted); potential safety concerns (infections, neutropenia) |
| Olokizu-mab | Antibody | KEGG Entry: D12487 | Kd ≈ 10 pM | [142] | Pros: high affinity, blocks IL-6 at site 3 preventing receptor complex formation<br><br>Cons: Not yet approved, high cost |
| SOMA-mer-SL1025 | DNA Aptamer | PDB 4N19 | Kd = 0.20 | [143] | Pros**:** high affinity, stable chemically modified DNA with slow off-rate. |
| | | | | | Cons: Requires chemical synthesis with modified nucleotides, susceptible to nuclease degradation, not yet used clinically |
| Clazaki-zumab | Antibody | KEGG Entry: D10312 | Kd ≈ 4 pM | [144] | Pros: High affinity, showed efficacy in clinical trials<br><br>Cons: Development ongoing, safety concerns (as therapeutic) |
| ZINC2997430 | Small molecule | ZINC ID: ZINC2997430 | Binding energy: −19.15 ± 4.04 kcal/mol | [145] | Pros: Small molecule, stable in 100 ns MD<br><br>Cons: No experimental Kd, only in silico evidence, off-target specificity untested, not commercially available |

| Growth Differentiation Factor-15 | | | | | |
|---|---|---|---|---|---|
| GFRAL | Protein | PDB 6WMW (GFRAL receptor w/ antibodies)<br><br>PDB 6Q2J (Complex w/ RET) | Kd = 8nM | [146] | Pros: natural high specificity, low immunogenicity<br><br>Cons: larger size, complex folding, not ideal for biosensors |
| APT2 | Aptamer | 5'-AGCAGCACAGAGGTCAGATG-N40-CCTATGCGTGCTACCGTGAA-3', wherein N is | KD = 1.12nM | [147] | Pros: Small DNA aptamer, high affinity, chemically stable, modifiable, G-quadruplex fold supports specificity |
| | | any A, T, G or C | | | Cons: no peer-reviewed validation, 3D structure unknown |

**Table 4.** Catalogue of experimental and in silico binders for the detection and targeting of key aging biomarkers: CRP, IL-6, IGF-1, GDF-15.